\documentstyle[preprint,tighten,aps,epsf]{revtex}

\begin{document}

\preprint{OU-HET 286}
\draft

\title{Lepton flavor violation at
       linear collider experiments in
        supersymmetric grand unified theories}
\author{Masahide Hirouchi and Minoru Tanaka%
        \thanks{E-mail: {\tt tanaka@phys.wani.osaka-u.ac.jp}}}
\address{Department of Physics,\\ 
         Graduate School of Science, Osaka University,\\
         Toyonaka, Osaka 560, Japan}
\date{\today}

\maketitle

\begin{abstract}
Lepton flavor violation at linear collider 
experiments is discussed. We show that detectable 
lepton flavor violation could occur through scalar 
lepton pair production and decay in the supersymmetric 
SU(5) grand unified theory in spite of the stringent present 
experimental constraints by rare process searches. Possible 
cross  sections about 40fb for an $e^+e^-$ collider and 
280fb for an $e^-e^-$ collider are illustrated.
\end{abstract}


\section{Introduction}
\label{INTRO}
Search for lepton flavor violation (LFV) is one of 
the most important ways to explore physics beyond 
the standard model, because lepton flavors are conserved 
individually in the standard model. In the standard model 
with supersymmetry (SUSY),  which is one of the most attractive
extensions of the standard model, 
LFV is allowed. The soft SUSY breaking masses 
of the scalar leptons (sleptons) 
do not have to conserve the lepton flavors in general. However,
the resulting LFV exceeds much the present 
experimental bounds such as the one from $\mu\rightarrow e\gamma$
search if an arbitrary, but consistent with the naturalness 
argument, set of slepton soft masses is allowed.
The universal soft mass scenario lead from supergravity \cite{NILLES}
is often assumed to avoid this large LFV (and the same problem in 
the scalar quark sector). In this scenario,
all the sleptons degenerate and we have no flavor mixing in 
the lepton and slepton sector.

However, this is not the whole story when grand unified
theories (GUTs) are considered at the same time \cite{HKR}. 
In SUSY GUTs with the universal soft breakings at Planck scale, 
the large top Yukawa coupling affects 
the third generation slepton mass
through renomalization group evolution from Planck scale to
GUT scale, since quark and lepton supermultiplets 
are unified into larger GUT multiplets. As a result, the sleptons
no longer degenerate and LFV is expected to take place.

Along this line, rates of $\mu\rightarrow e\gamma$ decay,
$\mu\rightarrow 3e$ decay, $\tau\rightarrow\mu\gamma$ decay and 
$\mu\rightarrow e$ conversion in nuclei have been estimated 
in the literature \cite{BHBHS,HMTY}. In the minimal SU(5) model, 
the $\mu\rightarrow e\gamma$ branching ratio has been found to be
typically one or two order below the present experimental 
upper bound ($4.9\times 10^{-11}$ \cite{BOLTON}).  
The other modes tend to give two or more order smaller 
values than experimental bounds. 
In the SO(10) model, the $\mu\rightarrow e\gamma$ amplitude
is enhanced by a factor $m_\tau/m_\mu$ due to its chiral structure
different from the SU(5) model, and thus the decay rate could be
the same order as or even larger than the experimental 
bound \cite{BHBHS}. 

Besides these rare processes that are caused by virtual
slepton exchange, it is possible to search for LFV through real
slepton production and its successive decay 
\cite{ACFH,ACFH-CP,KRASN-L,KRASN-H}. 
The most prominent qualitative difference 
between the virtual and the real processes
can be seen in behaviors of their amplitudes as the sleptons 
are getting to degenerate. The virtual process behaves as
$\Delta m^2/\bar m^2$, while the real process does as 
$\Delta m^2/(\bar m\Gamma)$ \cite{ACFH}, 
where $\Delta m^2$ is a slepton mass squared difference, 
$\bar m$ is the average mass of 
the sleptons and $\Gamma$ is the average slepton width. 
Because $\bar m\gg\Gamma$, we expect a good chance to
observe LFV even for relatively degenerate sleptons once their 
real production at collider experiments becomes possible.
The advantage of real production is maximized if 
$\Delta m^2/\bar m^2\ll 1 \ll\Delta m^2/(\bar m\Gamma)$ is 
realized. In the following, we show that it happens
in the minimal SUSY SU(5) model.

Another important point to have a realistic LFV cross section
of the real production and decay is necessity of relatively
large flavor mixing in the lepton-slepton sector. In the minimal
SUSY SU(5) model, the leptons and the down-type quarks have 
the same Yukawa couplings at the GUT scale. This means that LFV 
is essentially controlled by Cabibbo-Kobayashi-Maskawa (CKM) 
matrix which describes the quark mixing. 
Since CKM matrix is almost diagonal \cite{PDG}, 
LFV cross sections are suppressed by the small
off-diagonal elements of CKM matrix. It is shown below that
LFV cross sections at linear collider experiments are hopelessly
small in the minimal SUSY SU(5) model.

However, the minimal SUSY SU(5) model is no more than a calculable
example. In fact, it cannot describe the whole known fermion
masses and mixing \cite{BEREZ}. 
The above-mentioned equality of the down-type
quark Yukawa couplings and the leptonic ones leads to inconsistent
mass relation for the first and second generations, although
it gives the celebrated bottom to tau mass ratio. Once we extend
the model to overcome this insufficiency, the leptonic Yukawa
couplings are no longer necessary to be the same as the down-type 
quark Yukawa couplings. Thus, the leptonic mixing is independent 
of CKM matrix in general. 

In this paper, we show that LFV cross sections of charged slepton 
pair production and decay at linear collider experiments are
sizable if the lepton mixing has some appropriate structures. 
In Sec.\ \ref{FW}, we describe our framework based on 
SUSY SU(5) GUT more explicitly. 
Our numerical results on present experimental constraints and 
LFV cross sections are presented in Sec.\ \ref{NR}. Sec.\ \ref{CR} 
is devoted to concluding remarks.
 
\section{Framework}
\label{FW}
Apart from Yukawa sector, our working model is the minimal 
SUSY SU(5) GUT with the universal soft SUSY breaking terms 
at Planck scale, which detail is discussed in Ref.\ \cite{BHBHS} 
in the context of LFV.
The quarks and leptons are unified into 
three pairs of SU(5) chiral supermultiplets, 
$10$ ($T_i$) and $\bar 5$ ($\bar F_i$), 
with their superpartners, where $i=1,2,3$ is a generation index.
As for Higgs sector, we assume the minimal one, i.e.
$24$ ($\Sigma$), $5$ ($H$) and $\bar 5$ ($\bar H$).

In the minimal model, Yukawa superpotential is given by
\begin{equation}
 W_0=T_i f^{T}_{ij} T_j H + T_i f^{F}_{ij}\bar F_j\bar H,
\label{W0}
\end{equation}
where $f^T$ is the Yukawa coupling matrix which 
gives up-type quark masses, and $f^F$ is 
the one giving down-type quark
and charged lepton masses. We can take $f^T$ 
to be diagonal without loss of generality. 
Thus, the flavor mixing in the lepton
sector as well as the quark sector is governed 
by $f^F$ in the minimal model. As will be shown later, LFV 
cross sections of charged slepton pair production and decay are 
too small to be measured in this case.

However, as is mentioned in Sec.\ \ref{INTRO}, 
the above minimal model is known to predict incorrect mass 
relation for the first and the second generations.
To be realistic, it is natural to extend Eq.\ (\ref{W0}).
As a result, we expect that the leptonic Yukawa matrix is
different from the down-type one. For instance, this is realized
by introducing the following higher dimensional term
that might be induced by gravity \cite{EG}:
\begin{equation}
W_1= \frac{f^\prime_{ij}}{M_{\rm Planck}}
     \bar F_i^\alpha \Sigma_\alpha^\beta
      T_{j,\beta\gamma}\bar H^\gamma,
\label{W1}
\end{equation}
where the Greek indices are SU(5) ones.
Note that the effective Yukawa coupling
$f^\prime\langle\Sigma\rangle/M_{\rm Planck}\sim 
 f^\prime M_{\rm GUT}/M_{\rm Planck}$ 
could have the same
order of magnitude as $f^F$ due to the small masses of 
bottom and tau provided that $f^\prime\sim O(1)$
and $\tan\beta=\langle H\rangle/\langle\bar H\rangle$ is not 
too large.

In the following, we do not discuss specific extensions
like Eq.\ (\ref{W1}). Instead, we simply regard the leptonic 
mixing as independent from the quark mixing.

Lepton mass matrix at the weak scale, $M_e$, is diagonalized as
$\bar e_L M_e e_R=\bar \ell_L D_e\ell_R$
by unitary transformations $e_R=V_e \ell_R$ and $e_L=U_e\ell_L$,
where $M_e=U_e D_e V_e^\dagger$, 
$D_e={\rm diag.}(m_e,m_\mu,m_\tau)$,
$e_{R,L}$ denote the gauge eigen states, 
$\ell_{R,L}$ are the mass eigen states, 
and generation indices are suppressed. 
Making the same unitary transformations for
corresponding sleptons, we obtain the following $6\times 6$
charged slepton mass matrix:
\begin{equation}
\left(\tilde e_L^\dagger,\tilde e_R^\dagger\right)
\left(\begin{array}{cc}
       m_L^2 & m_{LR}^2 \\
       {m_{LR}^2}^\dagger & m_R^2
      \end{array}
\right)
\left(\begin{array}{c}
       \tilde e_L \\
       \tilde e_R
      \end{array}
\right),
\label{SLM}
\end{equation}
\[
m_L^2=\bar m_L^2\mbox{\boldmath $1$},\;
m_R^2=\bar m_R^2\mbox{\boldmath $1$}-
      V_e^\dagger\mbox{\boldmath $I$}V_e,\;
m_{LR}^2=-D_e(A_e\mbox{\boldmath $1$}-
              \frac{1}{3}V_e^\dagger\mbox{\boldmath $I$}^\prime V_e+
              \mu\tan\beta\,\mbox{\boldmath $1$}),
\]
\[
\mbox{\boldmath $I$}={\rm diag.}\,(0,0,I),\;
\mbox{\boldmath $I$}^\prime={\rm diag.}\,(0,0,I^\prime),
\]
where $\mbox{\boldmath $1$}$ is the $3\times 3$ unit matrix,
$\bar m_{L(R)}^2$ denotes the degenerate left(right)-handed 
charged slepton mass squared coming from the soft and electroweak 
breakings, $A_e$ is the universal soft breaking trilinear coupling
for slepton, $I$ denotes the shift of the soft breaking 
mass of the third generation
charged slepton during the renomalization group evolution
from $M_{\rm Planck}$ to $M_{\rm GUT}$ due to the large top
Yukawa coupling, and $I^\prime$
is the similar shift of the soft breaking trilinear coupling.
The renomalization group equations necessary to evaluate 
these quantities are found in Ref.\ \cite{BHBHS}.

Because of the degeneracy of the left-handed
slepton soft masses, $U_e$ does not appear in Eq.\ (\ref{SLM}).
Thus, LFV is controlled by a $3\times 3$ unitary matrix $V_e$. 
It turns out that $V_e$ contains only two parameters since,
as seen in Eq.\ (\ref{SLM}), the first and second generation 
right-handed slepton soft breaking parameters are the same. 
In addition, apparently no CP violating complex phase exists 
in $V_e$. In the following analysis, we take absolute values of 
(3,1) and (3,2) elements of $V_e$ as the parameters in $V_e$. 
We denote them as $|V_{31}|$ and $|V_{32}|$.

\section{Numerical results on LFV}
\label{NR}
By diagonalizing Eq.\ (\ref{SLM}) numerically, we calculate 
rates of several LFV processes as functions of $|V_{31}|$ and 
$|V_{32}|$. Masses and couplings of SUSY particles at the weak
scale are determined through the renomalization group equations
by giving the universal scalar mass ($m_0$), 
the GUT gaugino mass ($M_0$) and
the universal $A$ parameter ($A_0$) at Planck scale, 
in addition to sign of the supersymmetric Higgsino mass ($\mu$), 
$\tan\beta$, $|V_{31}|$ and $|V_{32}|$ at the weak scale. 
For an illustrative purpose, we take 
$m_0=100{\rm GeV}$, $M_0=150{\rm GeV}$ 
$A_0=0$, and  ${\rm sign}(\mu)=+1$. 
As for $\tan\beta$, results for $\tan\beta=3$ and $10$ 
are shown. Top quark mass is assumed to be 175GeV.
These input parameters are consistent with present 
experiments \cite{JANOT} other than LFV experiments 
discussed below for all possible values of $|V_{31}|$ 
and $|V_{32}|$. 
The lightest SUSY particle (LSP) is the lightest neutralino,
which is almost Bino, with mass around 63GeV depending
on $\tan\beta$. Mass spectrum of the six charged sleptons is
given as (150,163,163,182,182,183)GeV for $\tan\beta=3$ and
(149,164,164,183,183,190)GeV for $\tan\beta=10$. The precise 
values of the charged slepton masses depend on $V_e$ and
the above values are obtained for $V_e=\mbox{\boldmath $1$}$.
The lightest charged slepton, which production and decay
with LFV is discussed below, decays mostly into a LSP and 
a charged lepton. A typical decay width of the charged 
sleptons is about 0.5GeV for the above parameters.
Note that $\Delta m^2/\bar m^2\ll 1 \ll\Delta m^2/(\bar m\Gamma)$ 
is realized as mentioned in Sec.\ \ref{INTRO}.

Before discussing cross sections at linear collider experiments,
let us examine constraints on $|V_{31}|$ and 
$|V_{32}|$ from present LFV experiments. The constraint from 
${\rm Br}(\mu\rightarrow e\gamma)<4.9\times 10^{-11}$ \cite{BOLTON} 
is shown in Fig.\ \ref{PC}. We also show lines for $1\times 10^{-12}$
and $1\times 10^{-13}$. Fig.\ \ref{PC}(a) is for the case that 
$\tan\beta=3$, and Fig.\ \ref{PC}(b) is for $\tan\beta=10$.
For larger $\tan\beta$, the constraint is stronger because 
the $\mu\rightarrow e\gamma$ rate increases as $\tan\beta$ increases
\cite{BHBHS,HMTY}. We can see from Fig.\ \ref{PC} that 
$\mu\rightarrow e\gamma$ mainly gives a constraint on 
$|V_{31}V_{32}|$.
In Fig.\ \ref{PC}, we also show branching ratios of 
$\tau\rightarrow\mu\gamma$ and $\tau\rightarrow e\gamma$.
The present experimental upper bound for 
${\rm Br}(\tau\rightarrow\mu(e)\gamma)$ is 
$3.0(2.7)\times 10^{-6}$ \cite{CLEO}, and no constraint on 
the $|V_{31}|$--$|V_{32}|$ plane is obtained.
$\tau\rightarrow\mu(e)\gamma$
constrains $|V_{33}V_{32}|(|V_{33}V_{31}|)$ as is expected.

$\mu\rightarrow e$ conversion and $\mu\rightarrow 3e$ decay 
also give similar constrains on the $|V_{31}|$--$|V_{32}|$ plane
as $\mu\rightarrow e\gamma$. Since they tend to be weaker 
than the $\mu\rightarrow e\gamma$ constraint \cite{BHBHS,HMTY}, 
we do not discuss them for simplicity.

In Fig.\ \ref{CS3}, we show cross sections of LFV processes through 
pair production of the lightest charged sleptons at linear
collider experiments for the case that $\tan\beta=3$.
Fig.\ \ref{CS3}(a) shows the total cross section
of $e^+e^-_R\rightarrow\tilde\ell_1^+\tilde\ell_1^-
     \rightarrow \tau\mu+2{\rm LSPs}$, 
where $\tilde\ell_1^\pm$ denotes the lightest charged slepton. 
We assume $\sqrt{s}=500{\rm GeV}$ and 100\% right-handed
polarization of the electron beam. 
The $\mu\rightarrow e\gamma$ constraint
is also shown for comparison. We expect 40fb as the maximal
cross section in spite of the strong constraint because of 
different dependence on $V_e$ from $\mu\rightarrow e\gamma$. 
This process depends on $V_e$ mainly 
through combinations of $|V_{33}V_{32}|$ and 
$|V_{33}V_{32}V_{31}^2|$ corresponding to the s- and t- channel 
diagrams respectively. The allowed maximal LFV cross section is
obtained in the case that $|V_{31}|\ll |V_{32}|\simeq|V_{33}|$. 
Note that the LFV cross section is of
order 0.1fb if $V_e$ has a similar structure to CKM matrix
as in the minimal SU(5) model.

Fig.\ \ref{CS3}(b), (c) and (d) show the total cross cross sections
of $e^+e^-_R\rightarrow\tilde\ell_1^+\tilde\ell_1^-
     \rightarrow \tau e+2{\rm LSPs}$,
$e^-_Re^-_R\rightarrow\tilde\ell_1^-\tilde\ell_1^-
     \rightarrow \tau\tau+2{\rm LSPs}$, and
$e^-_Re^-_R\rightarrow\tilde\ell_1^-\tilde\ell_1^-
     \rightarrow \tau e+2{\rm LSPs}$, respectively.
The same $\sqrt{s}$ and beam polarization are assumed as 
Fig.\ \ref{CS3}(a). The maximal values of cross sections are about 
150fb, 80fb, and 280fb respectively. These values are realized in
the case that $|V_{32}|\ll |V_{31}|\sim |V_{33}|$.

Other lepton flavor violating combinations in the final state 
charged lepton pair give less interesting cross sections 
for both $e^+e^-_R$ and $e^-_Re^-_R$ collisions.

Fig.\ \ref{CS10} shows the same quantities as Fig.\ \ref{CS3}, 
but for $\tan\beta=10$. Comparing Fig.\ \ref{CS10} with 
Fig.\ \ref{CS3}, we find that the LFV cross sections are smaller 
for larger $\tan\beta$. This means that these processes are 
complimentary to $\mu\rightarrow e\gamma$ which is enhanced
for larger $\tan\beta$.

The reduction of the LFV cross sections for larger $\tan\beta$
is due to large left-right mixing of scalar tau leptons. To see
this, it is enough to consider the following $3\times 3$ 
submatrix of Eq.\ (\ref{SLM}):
\begin{eqnarray}
& &\!\!\!M^2=\nonumber\\
& &\!\!\!\left(\begin{array}{ccc}
              (m_L^2)_{33} & (m_{LR}^2)_{32} & (m_{LR}^2)_{33}\\
              {\rm *} & (m_R^2)_{22} & (m_R^2)_{23}\\
              {\rm *} &{\rm *} & (m_R^2)_{33}
	      \end{array}
        \right)
    \simeq \left(\begin{array}{ccc}
              \bar m_L^2 & 0 & -m_\tau\mu\tan\beta\\
              {\rm *} & \bar m_R^2-IV_{32}^2 & -IV_{32}V_{33}\\
              {\rm *} & {\rm *} & \bar m_R^2-IV_{33}^2
              \end{array}
      \right),
\end{eqnarray}
where we neglect terms other than the one 
proportional to $m_\tau\tan\beta$ in $m_{LR}^2$. 
Since we are considering a large $\tan\beta$ case, 
the left-right mixing angle of the scalar tau leptons is 
almost 45 degree. By making the 45 degree rotation 
in the 1--3 plane of $M^2$, we obtain a
matrix closer to a diagonal form:
\begin{eqnarray}
& &O_0^T M^2 O_0=\nonumber\\
& &\left(
 \begin{array}{ccc}
  \!\!\!\frac{\bar m_L^2+\bar m_R^2-IV_{33}^2}{2}\!+\!
  m_\tau|\mu|\tan\beta &
 -\frac{\mu}{\sqrt{2}|\mu|}IV_{32}V_{33} &
 \frac{\mu}{2|\mu|}(\bar m_L^2-\bar m_R^2+IV_{33}^2)\\
 {\rm *} & \bar m_R^2-IV_{32}^2 & IV_{32}V_{33}/\sqrt{2} \\
 {\rm *} & {\rm *} & \!\!\!\!\!\!
               \frac{\bar m_L^2+\bar m_R^2-IV_{33}^2}{2}\!-\!
                      m_\tau|\mu|\tan\beta 
 \end{array}
\!\!\!\right),
\label{RSLM45}
\end{eqnarray}
where $O_0$ is the 45 degree rotation matrix:
\begin{equation}
O_0=
\left(\begin{array}{ccc}
       \frac{1}{\sqrt{2}} & 0 & \frac{\mu}{|\mu|\sqrt{2}}\\
                    0 & 1 & 0 \\
       -\frac{\mu}{|\mu|\sqrt{2}} & 0 & \frac{1}{\sqrt{2}}
                    \end{array}
\right).
\end{equation}
It is legitimate to diagonalize Eq.\ (\ref{RSLM45}) perturbatively 
in order to see qualitative behavior of the slepton flavor mixing. 
As a result, we find that LFV related off-diagonal elements of
the slepton mixing matrix are approximately proportional to 
\begin{equation}
\sim\frac{I}{m_\tau|\mu|\tan\beta}
\end{equation}
in the large $\tan\beta$ case. Thus, the LFV cross sections are
suppressed as $\tan\beta$ becomes large.

\section{Concluding remarks}
\label{CR}
Before concluding we discuss some background issues.
Possible extensions of our calculation are also 
discussed here.

Our LFV signals are $\tau^\pm\ell^\mp+{\rm missing}$ where 
$\ell$ denotes $e$ or $\mu$ for the $e^+e^-$ collision,
and $\tau^-\ell^-+{\rm missing}$ with $\ell=e$ or $\tau$ for 
the $e^-e^-$ collision. The tau leptons are identified with 
their hadronic decays. The pure leptonic decay modes can also 
be used with impact parameter analysis in principle. 
A CCD pixel vertex detector proposed for linear collider
experiments has a typical resolution better than 10$\mu$m \cite{DJ}, 
while $c\tau$ of tau lepton is about 90$\mu$m \cite{PDG}.

The most serious standard model background in 
$e^+e^-$ collision is the one from $W$ boson pair production. 
A leptonic $W$ pair decay $WW\rightarrow\tau\ell\nu\nu$ 
($\ell=e,\mu$) is a background event. 
$WW\rightarrow\tau\tau\nu\nu$ followed 
by a pure leptonic decay of one of the $\tau$'s can also 
be a background if the above-mentioned impact parameter analysis
is not effective. However, these $WW$ backgrounds are reduced 
by employing a right-handed electron beam as we did in the above 
calculation. Appropriate kinematical cuts are also useful.
With these procedures, the background cross section is
reduced to less than $O(1)$fb with a reasonable efficiency of 
about $30\sim 50\%$ for the signal \cite{BV,NOJIRI}. 
$e\bar\nu W$ and $eeWW$ could also be backgrounds of 
about 1fb \cite{BV}. 

$ZZ$ $Zh$, and $ee\tau\tau$ could also be backgrounds. 
They can be reduced to $O(1)$fb or less by appropriate 
selection cuts \cite{BV,NOJIRI}. Moreover, the impact parameter 
analysis could reduce these modes to negligible levels.

As for the $e^-e^-$ collision, our LFV signals are essentially 
free from backgrounds. In particular, the right-handed 
beams reduce $e^-\nu W^-$ mode to a negligible level without any
selection cuts \cite{COR}.

Production of other superparticles could also be background.
In particular, heavier slepton production makes LFV signals
more complicated. To avoid it, we can choose such a $\sqrt{s}$
that only a pair of the lightest charged sleptons is created. 
Although the cross sections decrease, especially for the s-channel 
contribution in the $e^+e^-$ collision, we still have sizable 
cross sections both for the $e^+e^-$ and $e^-e^-$ collisions.

Our observations in the present work can also be applied to
other SUSY GUTs qualitatively. For instance, the LFV cross 
sections are expected to be sizable in the SO(10) model. 
In this model, the left-handed sleptons, as well as the right-handed
ones, cause LFV since the left-handed lepton supermultiplet of 
the third generation is unified into the same GUT multiplet as the 
top quark. Then, as is mentioned in Sec.\ \ref{INTRO},
the $\mu\rightarrow e\gamma$ rate is enhanced by 
a factor $\sim m_\tau/m_\mu$ because of the chirality-flip nature
of this process. While $\tau\rightarrow\mu(e)\gamma$
is not enhanced. Although the detail depends on the model, 
especially its Yukawa superpotential, we expect a stronger 
constraint from $\mu\rightarrow e\gamma$ and similar constraints
from $\tau\rightarrow\mu(e)\gamma$ compared with the
SU(5) model. As can be seen in Fig.\ \ref{CS3} and \ref{CS10}, 
a stronger $\mu\rightarrow e\gamma$ constraint alone does not 
exclude sizable LFV cross sections.

We also expect sizable LFV cross sections at a muon collider
experiment \cite{CHENG}. The s-channel amplitude in 
$\mu^+\mu^-$ collision is the same as $e^+e^-$ case. 
While the t-channel one has different dependence on $V_e$.
The lower initial state radiation of muon collider would
make the threshold operation mentioned above more effective.

In conclusion, we have shown that the LFV phenomena through
slepton pair production and decay could be detectable at
linear collider experiments in the SUSY SU(5) GUT. Treating 
the lepton mixing matrix as a set of parameters independent
of CKM matrix, we have discussed the constraints on it from 
the present experiments and calculated LFV cross sections. 
In spite of the stringent constraint from
$\mu\rightarrow e\gamma$, some of LFV processes which
have tau lepton(s) in the final state could have sizable cross
sections at future linear collider experiments.

\begin{figure}
\caption{Present experimental constraint on 
         $|V_{31}|$ and $|V_{32}|$ from 
         $\mu\rightarrow e\gamma$ search: (a) for $\tan\beta=3$ and
         (b) for $\tan\beta=10$. 
         Br($\tau\rightarrow\mu(e)\gamma$) is also shown.}
\label{PC}
\end{figure}

\begin{figure}
\caption{LFV cross sections at $\sqrt{s}=500{\rm GeV}$ for 
         $\tan\beta=3$. The $\mu\rightarrow e\gamma$ constraint
         is also shown for comparison.}
\label{CS3}
\end{figure}

\begin{figure}
\caption{The same as Fig.\ \ref{CS3} except for 
         $\tan\beta=10$.}
\label{CS10}
\end{figure}

\newpage
\hspace*{-1.5cm}
\epsfbox[70 495 535 738]{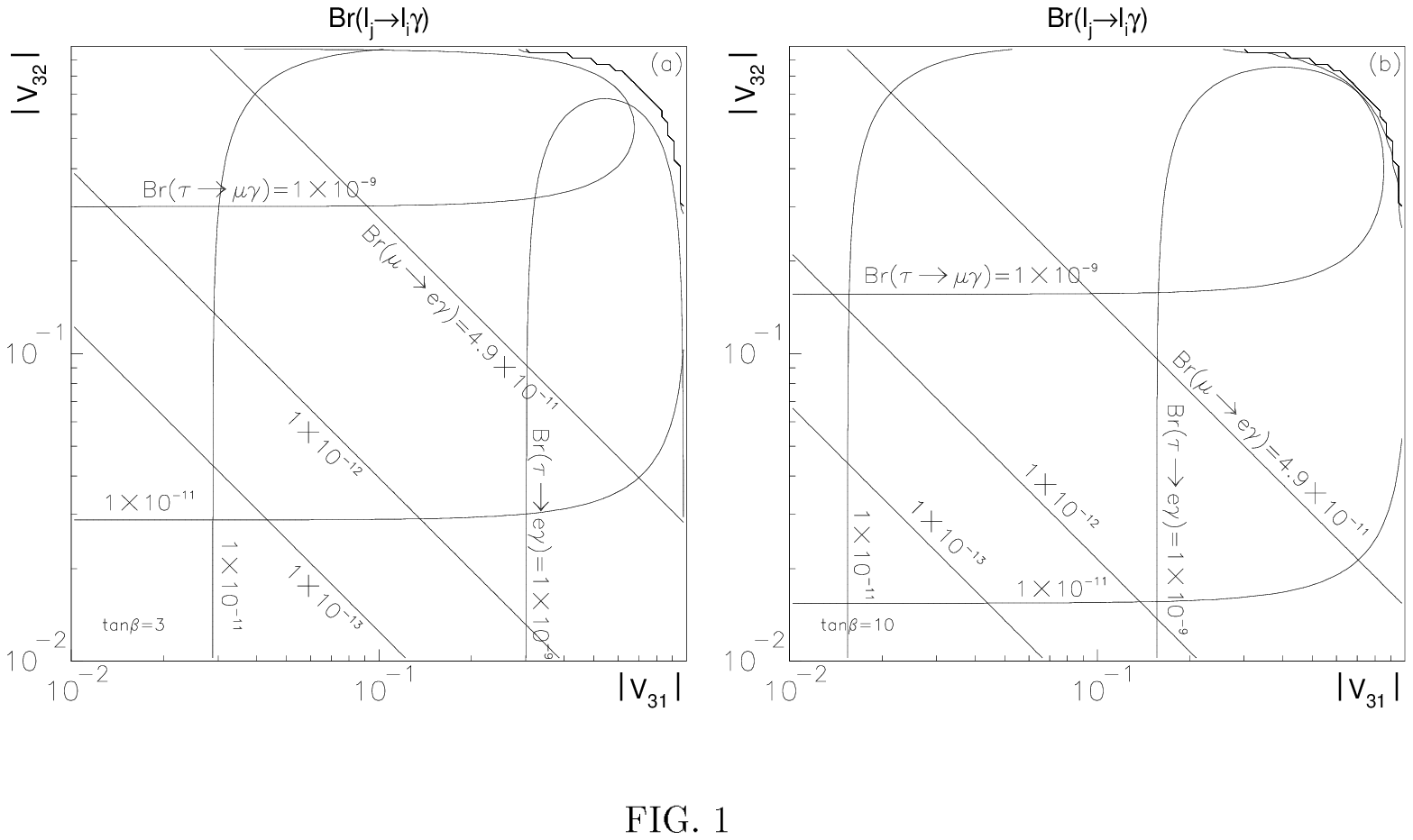}

\hspace*{-1.5cm}
\epsfbox[70 260 535 738]{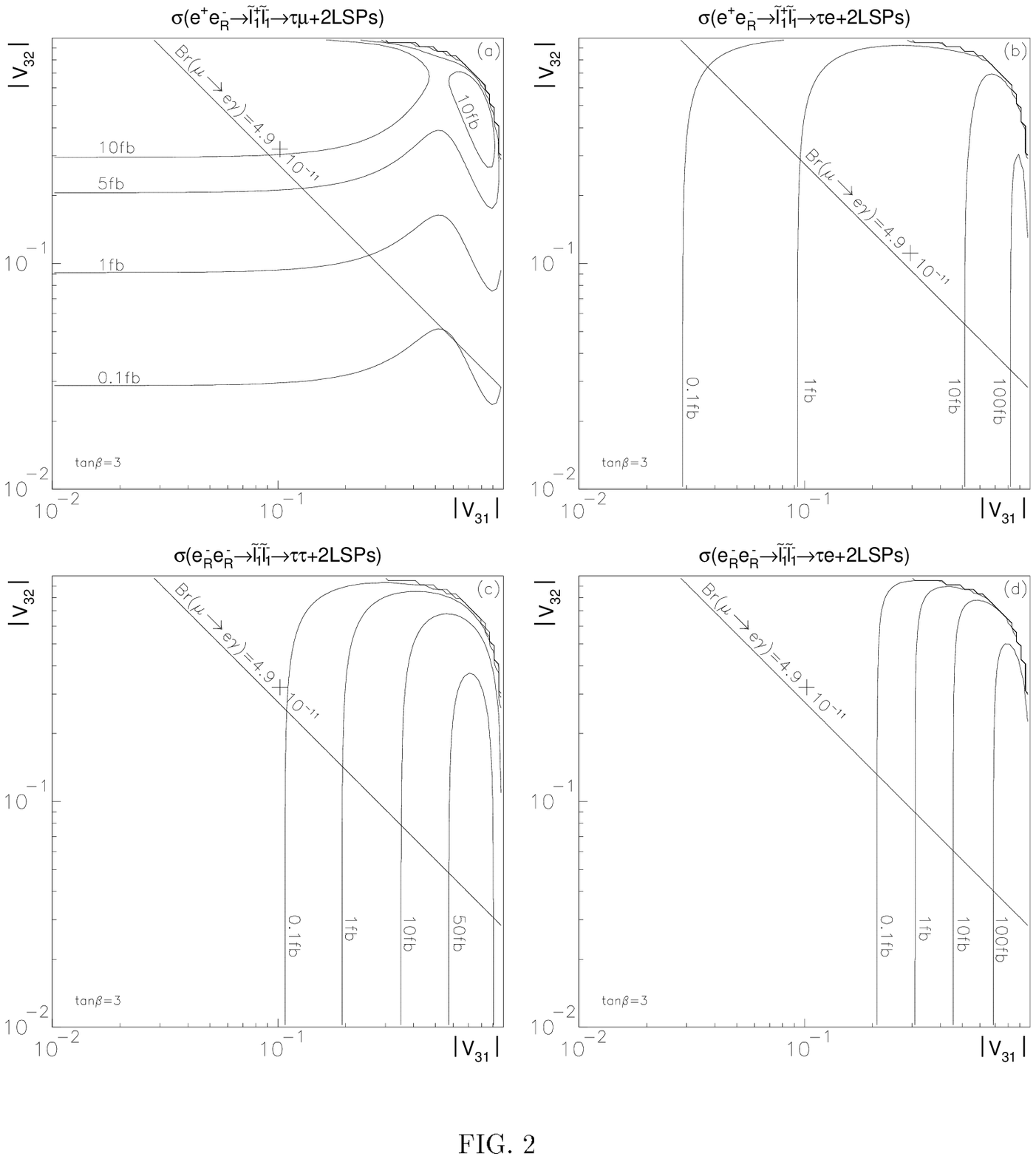}

\hspace*{-1.5cm}
\epsfbox[70 260 535 738]{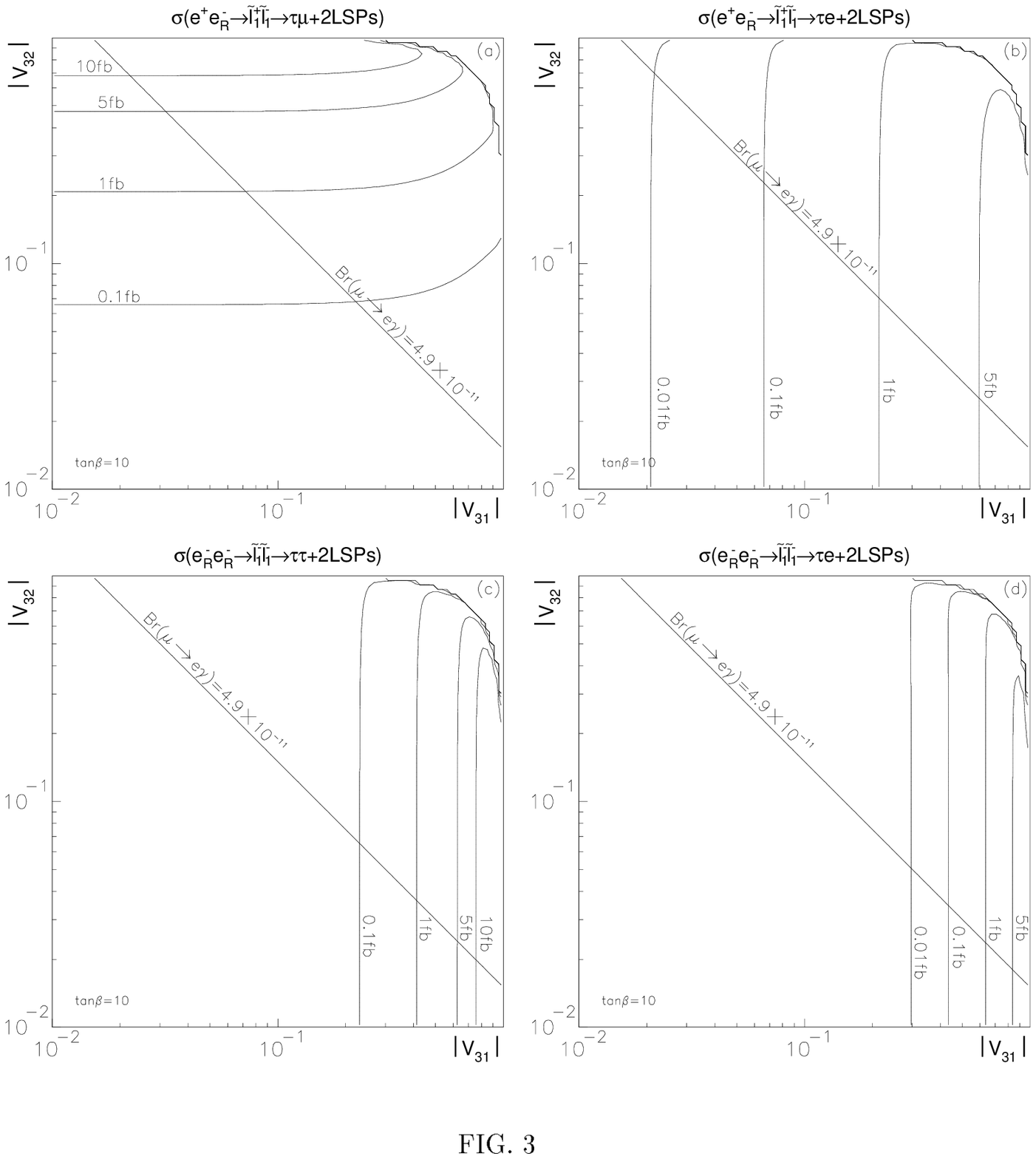}

\end{document}